\documentclass[twocolumn,amsmath,amssymb,nofootinbib,superscriptaddress,longbibliography]{revtex4-1}

\usepackage{graphicx}
\usepackage{amsmath}
\usepackage{amssymb}
\usepackage{xcolor}
\usepackage{hyperref}
\hypersetup{pdfstartview={FitH},pdfpagemode={UseNone}, breaklinks=true,
            colorlinks=true,bookmarks=false,linkcolor=blue, citecolor=blue, urlcolor=blue,
             bookmarksopen=true, pdfnewwindow=true}
\usepackage[all]{hypcap}
\usepackage[english]{babel}
\usepackage{physics}

\begin{document}

\preprint{AIP/123-QED}

\title{Energy harvesting coil for circularly polarized fields in magnetic resonance imaging}

\author{P. Seregin}
%\email{pavel.seregin@metalab.ifmo.ru}
\thanks{These two authors contributed equally}
\affiliation{Department of Physics and Engineering, ITMO University, Saint Petersburg, Russia}

\author{O. Burmistrov}
\thanks{These two authors contributed equally}
\affiliation{Department of Physics and Engineering, ITMO University, Saint Petersburg, Russia}

\author{G. Solomakha}
\affiliation{Department of Physics and Engineering, ITMO University, Saint Petersburg, Russia}
\author{E. Kretov}
\affiliation{Department of Physics and Engineering, ITMO University, Saint Petersburg, Russia}
\author{N. Olekhno}
\affiliation{Department of Physics and Engineering, ITMO University, Saint Petersburg, Russia}
\author{A. Slobozhanyuk}
 \email{a.slobozhanyuk@metalab.ifmo.ru}
\affiliation{Department of Physics and Engineering, ITMO University, Saint Petersburg, Russia}

\date{\today}

\begin{abstract}

Specialized  RF coils and sensors placed inside the magnetic resonance imaging (MRI) scanner considerably extend its functionality. However, since cable connected  in-bore devices have several disadvantages compared to wireless ones, the latter currently undergo active development. One of the promising concepts in wireless MRI coils, energy harvesting, relies on energy carried by the RF MRI field without the need for additional transmitters as in common wireless power transfer. In this letter, we propose a compact harvesting coil design based on the combination of the loop and butterfly coils that allows energy harvesting of a circularly-polarized field. By performing numerical simulations and experiments with commonly used Siemens Espree and Avanto 1.5 Tesla MRI scanners, we have demonstrated that the proposed approach is safe, efficient, does not decrease the quality of MRI images, and allows doubling the harvested voltage in contrast with linearly polarized setup. 

\end{abstract}

\maketitle

%________________________Introduction____________________________
\section{Introduction}

\begin{figure}[b]
    \centering
    \includegraphics[width=8cm]{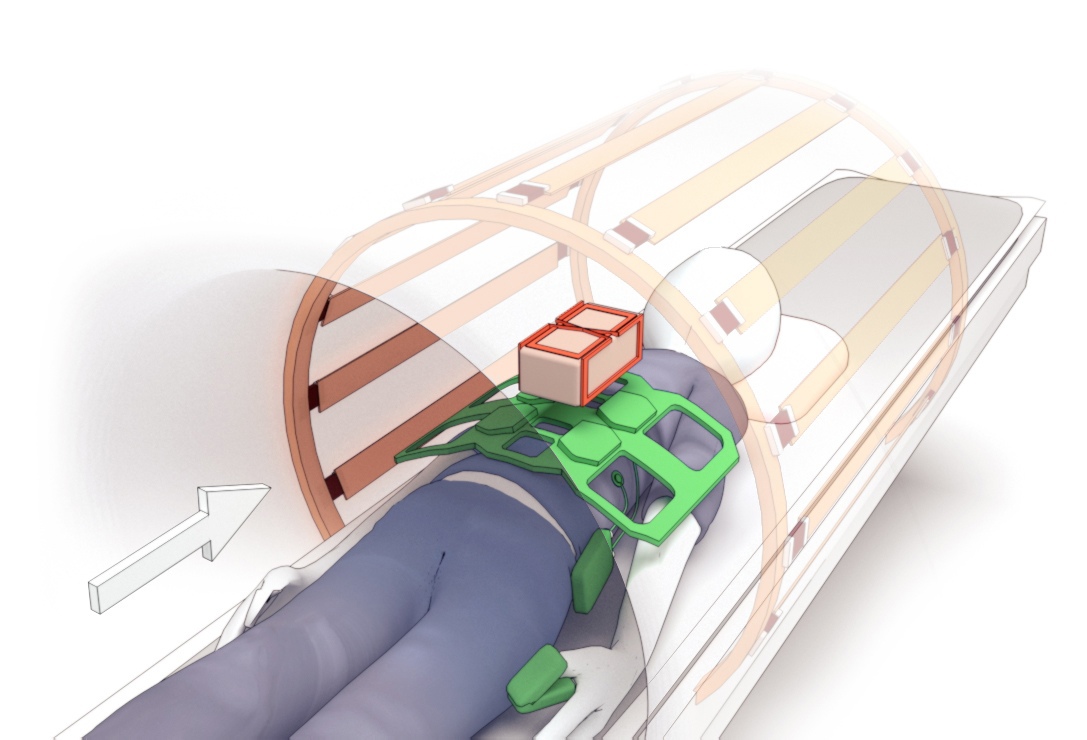}
    \caption{Artist's view of the harvesting setup. The patient located in the MRI scanner (shown with grey) with a large constant magnetic field $B_0$ applied. $B_0$ is depicted by white arrow. The harvesting coil (shown with red) is placed on top of the patient and powers up in-bore devices (shown with green) by converting the energy of RF field $B_1$. Examples of such equipment include local coils, heart activity monitors, and pulse sensors.}
    \label{fig:Harvesting}
\end{figure}

Magnetic resonance imaging (MRI) has rendered as one of the key methods in medical diagnostics. To further extend its functionality, various in-bore devices are applied, including different flexible body receive coils and sensors. However, these devices require energy supply, and powering them with wire lines can result in imaging artifacts or patient discomfort~\cite{Nohava2020}. Thus, various wireless alternatives are actively studied, including wireless power transfer (WPT) ~\cite{byron2017rf,song2020multi,byron2014mri, byron2019, ganti2019,song2016wireless_1,song2016wireless_2}, wireless MRI coil~\cite{ivanov2020coupled} and MRI energy harvesting ~\cite{Hofflin2013}. While the WPT requires the presence of additional transmitting coils, the harvesting relies on the conversion of electromagnetic fields that are already presented in the region of interest, such as $B_1$ radio-frequency (RF) field, Fig.~\ref{fig:Harvesting}, thus not requiring any modifications of the MRI scanner itself.

RF harvesting setups converting the electromagnetic energy present in the surrounding environment have been readily demonstrated considering GSM~\cite{bakir2018tunable, le2008efficient, oh2019low, devi2012design, uzun2015design}, RFID~\cite{huang2013adaptive, shahmohammadi2016high}, Wi-Fi~\cite{khansalee2015high, munir2016optimization}, and LTE~\cite{fantuzzi2016large} signals. The power level of signals in the environment for the conventional wireless communication bands has a typical power density  ~\cite{li2020progress, pinuela2013ambient, hemour2014radio, kim2014ambient} ranging from $2~\mu {\rm W}/{\rm m}^2$ to $10~{\rm mW}/{\rm m}^2$, which is much lower than that in MRI scanner. However, in contrast to such high-frequency applications, there are several differences when considering the application of harvesting concept in MRI: (i) MRI harvesting should work at $63~{\rm MHz}$ for $1.5~{\rm T}$ MRI and $128~{\rm MHz}$ for $3~{\rm T}$ MRI, which means near-fields region due to small MRI bore areas (usual it is approximately 70 cm in diameter); (ii) MRI scanners generate RF-fields in the form of short high-power pulses up to $10$~ms long; (iii) There is a large constant magnetic field $B_{0}$ present, which requires the use of non-magnetic materials; (iv) Currents flowing in the harvesting circuit should not distort the magnetic field $B_{1}$ and $B_{0}$ which can result in the image quality decrease; (v) Since MRI is a method of medical visualization, strict safety requirements should be met, including the limitation of specific absorption rate (SAR).

The general principles of energy harvesting and its application in MRI are illustrated in Fig.~\ref{fig:Harvesting}. Inside an MRI scanner, a large constant magnetic field $B_0$ is created, which we assume to be lower than $3$~T as a body coil is needed for the energy harvesting to function, while usually, MRI setups with $B_0>3$~T do not feature a body coil due to its low efficiency and strong field homogeneity in the sample~\cite{Vaughan_2009}. A body coil used for the transmission, and a local coil for the reception of echo signals from protons ~\cite{Gruber_2018}. In the process of MRI scanning, additional pulse gradient fields $G_x(t)$, $G_y(t)$, and $G_z(t)$ are created in order to obtain a spatial resolution ~\cite{haackeMRI}, which have typical values of ${\rm mT/m}$~\cite{haackeMRI} and produce up to $1$~MW of peak power~\cite{Gradient_paper}. However, a realization of an effective coil for their conversion is very challenging~\cite{Hofflin2013} since a large coil is necessary to capture spatial field variations. Finally, an RF field $B_1$ is created for the excitation of the nuclear magnetic resonance at the Larmor frequency~\cite{haackeMRI}. This field is {\it circularly} polarized and has peak pulse power of $10...15$~kW~\cite{cpc_amps}, which is potentially enough to provide DC supply for in-bore electronics, including circuits of local coils, as well as electro-cardio-graphic (ECG) gating, respiratory (chest movement tracking), and pulse (heart rate capturing) sensors of physiological signals, or external triggering source~\cite{Siemens_Pr_Guide}. Moreover, energy harvesting coils for RF field $B_1$ can be effectively implemented as they do not require a large size or multi-turn design for functioning in contrast to schemes using gradient coils.

This resulted in several realizations of harvesting setups in MRI~\cite{riffe2007, Hofflin2013}. However, all of the mentioned setups rely on converting linearly polarized components of the field $B_1$, thus limiting the harvesting efficiency. In the present letter, we extend the proposed concepts and develop a harvesting coil capable of capturing a circular polarization in order to enhance the harvesting efficiency.

%____________________________Design___________________________
\section{Coil design for circularly polarized wave harvesting}

\begin{figure}[t]
   %% \centering
    \includegraphics[width=8cm]{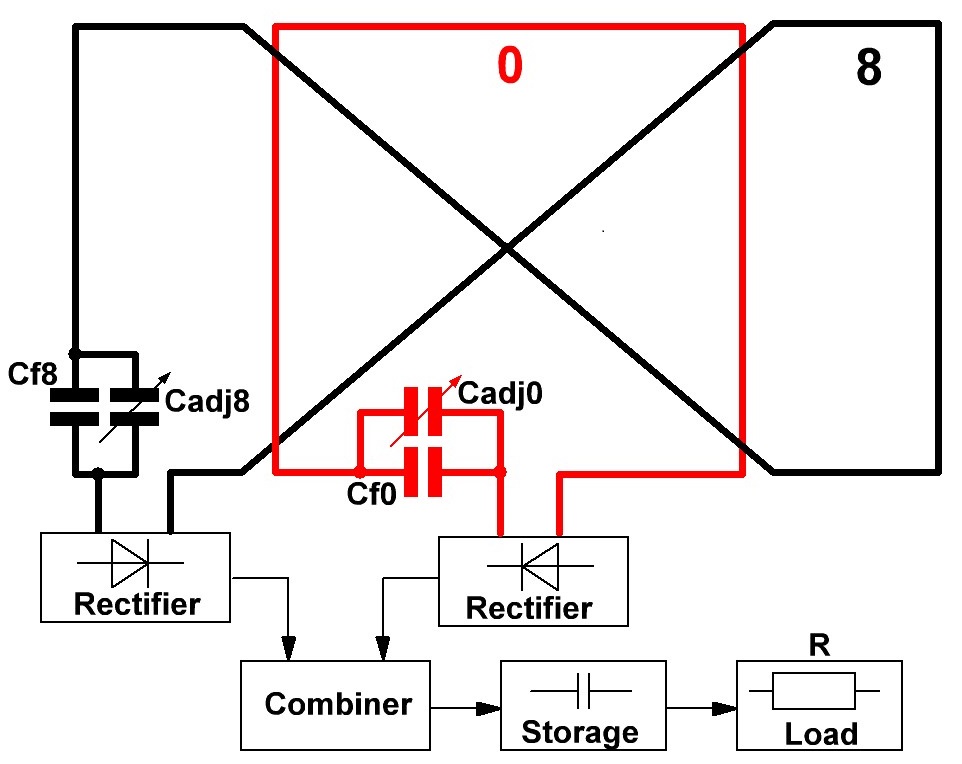}
    \caption{Schematics of the developed circularly polarized field harvesting setup consisting of two coils, one in the shape of a loop (shown in red) and another one in the shape of a butterfly (shown in black). }
    \label{fig:Circuit}
\end{figure}

The proposed coil concept illustrated in Fig.~\ref{fig:Circuit} includes two coils. One having the shape of a circle ($0$-shape) and another one with the shape of a butterfly ($8$-shape). Two coils are required to convert orthogonal linear parts of the circular polarization of the field $B_1$, with one linear component being captured by each coil. Considering several coil designs discussed in Supplementary Materials, we obtained the best parameters for the presented configuration. The parameters taken into account include: (i) Decoupling circuitry (additional inductors are needed for planar designs realized within a single printed circuit board, while for non-planar schemes, purely-geometrical decoupling can be applied); (ii) Conversion efficiency; (iii) Simplicity of implementation. The harvesting setup also includes two tuning capacitors, one for each coil, that are needed to precisely adjust the coil resonance to the Larmor frequency. The currents generated in coils then are rectified by voltage doublers to enhance the conversion voltage. 

The value of DC current after rectifiers can reach up to relatively high values. This can lead to the emergence of an effective magnetic field disturbing the $B_0$ field. Since MRI requires an extreme homogeneity of $B_0$ field within the imaging volume, such a perturbation can result in considerable imaging artifacts~\cite{twieg2015active}. To eliminate such effects, we implement all current-carrying lines in a combiner with differential pairs. Further details are given in Supplementary Materials.

The size of coil elements defines voltages obtained with harvesting setup: larger sizes result in larger voltages and harvested power, respectively. However, large sizes of harvesting coil may lead to the distortion of RF field $B_1$ in the vicinity of coil and decrease the patient's comfort. In experiments and numerical simulations, we use the following coil parameters: length $L=110$~mm, width $W=65$~mm, and height $H=35$~mm, Fig.~\ref{fig:Experiment}(a). Each tuning capacitor is realized as two capacitors arranged in parallel, a fixed capacitor $C_{f}$ and adjustable capacitor $C_{adj}$. For the considered coil size, the following values are chosen: $C_f=4~p{\rm F}$ and $C_{adj}=2...12~p{\rm F}$. Storage capacitor $C$ consists of several capacitors arranged in parallel with a total capacitance of $C=470~\mu{\rm F}$. Two rectifiers were realized using Schottky diodes. Finally, the rectified currents from two coils are combined using RF-traps and charge the storage capacitor which powers the load 200 Ohm, which is equivalent to a load of four low noise amplifiers from a standard Siemens body flex coil. 

To perform experimental studies of the developed harvesting coils with a real MRI scanners, we use the following setup illustrated in Fig.~\ref{fig:Experiment}(b). The harvesting coil was placed at the top of a standard calibrating phantom in form of a cylindrical bottle with dimensions $520 \times 440 \times 340$~mm filled with 1,24g ${\rm NiSO}_{4} \times 6{\rm H}_{2}{\rm O}$/2,62g NaCl per 1000~ml distilled water. The phantom is placed upon standard pads, and together with coil located inside a cylindrical Siemens loaded body phantom MRI 5750307 filled with the same liquid. All the mentioned elements are placed on the patient table allowing to move them inside the MRI scanner.

%____________________________Experiment____________________________
\section{Experimental results}

\begin{figure}[t]
    \centering
    \includegraphics[width=7cm]{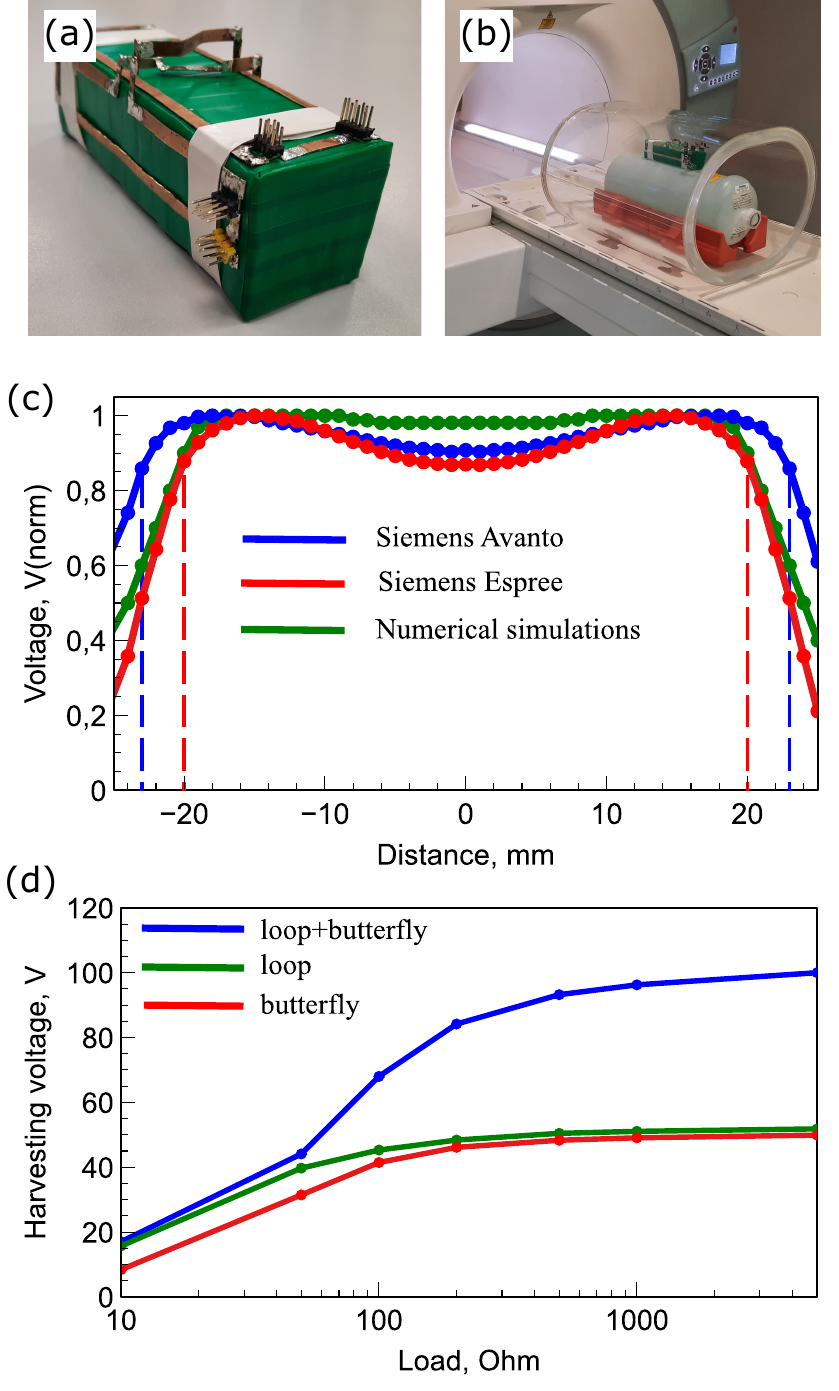}
    \caption{(a) Photo of the experimentally realized coil for circularly polarized field harvesting. (b) Experimental setup including an MRI scanner, body phantom consisting of a bottle and Siemens loading body phantom, and the harvesting coil atop of the phantom. (c) Normalized harvesting voltage value versus coil position. The region between dashed lines is suitable for harvesting. (d) Harvesting voltage value versus load resistance}
    \label{fig:Experiment}
\end{figure}

We start with measuring output voltage $V_{ind}$ induced at the load $R_{load}$ with RF power applied to the body coil. We study the dependence of voltage $V_{ind}$ on the distance between the harvesting coil position and the isocenter of MRI bore magnet $z$. To change the location of harvesting coil, we moved it manually while keeping the positions of the table and phantom unchanged. Such a procedure is chosen to prevent a change of the transmitting body coil impedance. The output voltage was measured using Owon XDS3202A 14-bit oscilloscope attached by a coaxial cable with multiple RF-traps to prevent cable antenna effect. We have measured the output voltage with a single excitation pulse during transmit, as RF harvesting should provide voltage to the local coil circuitry during each phase encoding step. To proceed, we applied a manual frequency adjustment procedure employing three separate pulses. However, only the first one was used to measure the output harvesting voltage. The sets of voltage measurements have been performed utilizing Siemens Avanto (red curve), Siemens Espree (blue curve) $1.5$~T MRI systems and numerical simulations with CST Microwave Studio 2020 (green). The resulting voltage dependence $V(z)$ is shown in Fig.~\ref{fig:Experiment}(c). We use geometric dimensions equivalent Espree BC for the CST numerical model.The graphs show that the width of the curves is slightly different due to the fact that the length of body coils is different. The longer the BC length, the larger the working area for harvesting can be obtained. As seen from the figure, the voltage first slowly increases when the harvesting coil is moved from the isocenter, reaching its maximum at $z \approx 14$~cm for Espree scanner and $z \approx 18$~cm for Avanto scanner. However, the voltage rapidly drops at further increase of the distance $z$, and the harvesting setup becomes inefficient. To estimate the region in which the harvesting setup maintains stable functionality, we select a $15\%$ drop in the output voltage as a criterion. Then, as showed by dashed lines in Fig.~\ref{fig:Experiment}(c), the realized harvesting coil can be freely moved within $\pm 20$~cm range for Espree setup and within $\pm 22.5$~cm range for Avanto setup without considerable change in the output voltage. This is larger than the typical MRI coil size, thus offering a considerable degree of freedom in placing the harvesting coil. The measurements show that it is possible to increase output voltage almost two times at the harvesting output with circularly polarized RF-coil. Still, the overall efficiency depends on the value of the load. Further details are given in Supplementary Materials.

\begin{figure*}[t]
    \centering
    \includegraphics[width=17cm]{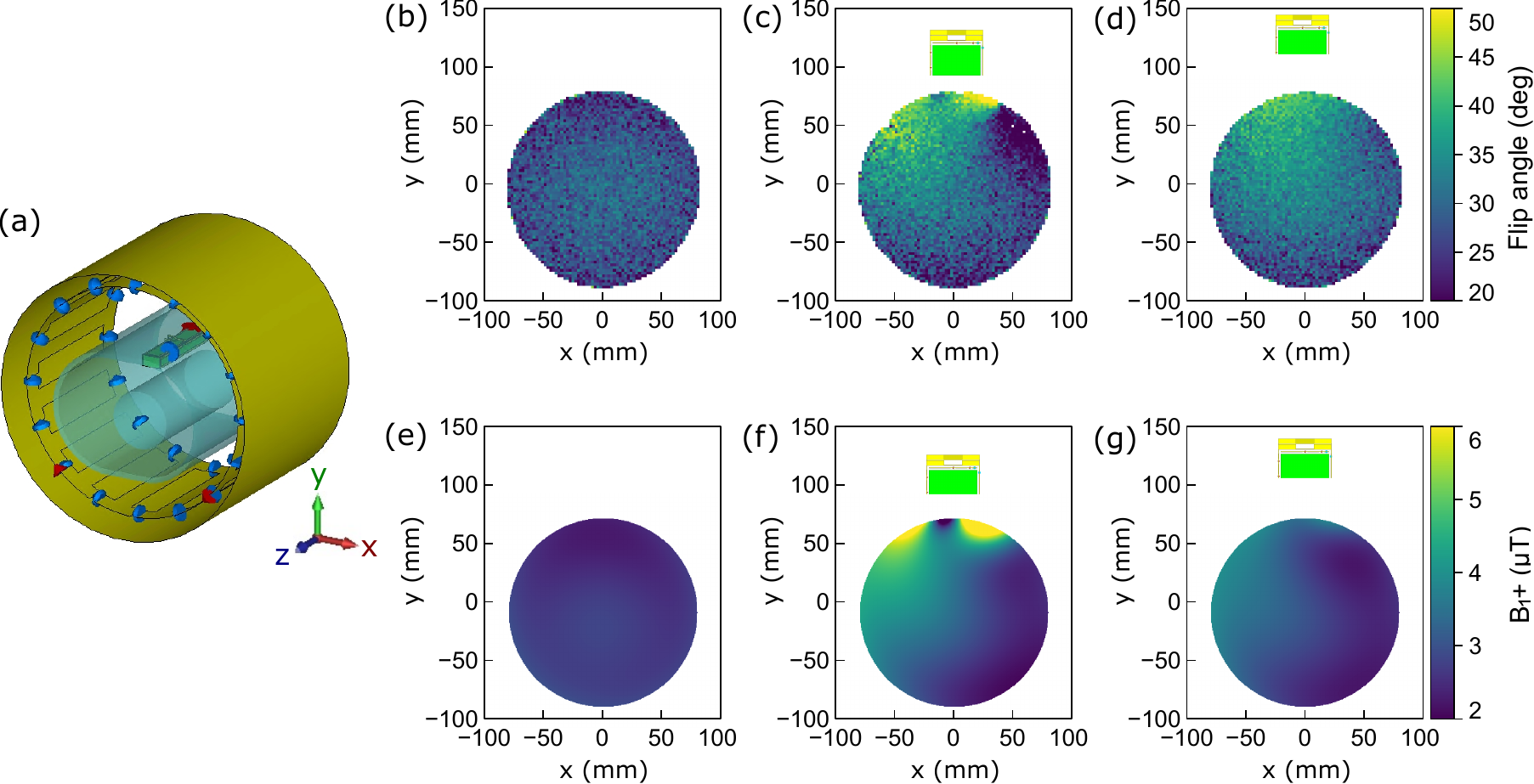}
    \caption{(a) The numerical model implemented in CST Microwave Studio. Blue cylinder and frame correspond to the water-filled body phantom. BC and scanner edge are shown with yellow, while blue disks denote capacitors placed in birdcage coil. The harvesting coil is placed atop the phantom. Red arrows show the location of ports used in numerical simulation. Maps of the RF magnetic fields $B_1$ in the $(xy)$-plane for $Z=0$ obtained numerically (e-g) and experimentally (b-d) for the phantom without harvesting coil (b,e), with harvesting coil located at the height $h=1$~cm above the phantom (c,f), and with harvesting coil located at the height $h=5$~cm above the phantom (d,g).}
    \label{fig:B1_MAP}
\end{figure*}

Next, we study the RF field $B_1$ inside the body phantom for different locations of the harvesting coil to ensure that it does not impact the MRI functionality. The axial cross-section of $B_1$ field in $(xy)$-plane at the isocenter $z=0$ for Espree MRI scanner with a body phantom is shown in Fig.~\ref{fig:B1_MAP}(b). In such a case, no harvesting coil is present, and the magnetic field appears nearly homogeneous up to some noise which is caused mostly by the pulse sequence chosen for $B_1$ measurement within the applied technique~\cite{insko1993mapping}. Then, if the harvesting coil is introduced in the close vicinity of the phantom's surface at the distance of $1$~cm, the field $B_1$ considerably changes (see Fig.~\ref{fig:B1_MAP}(c)). In particular, the presence of coil leads to the emergence of inhomogeneity at the right side of the phantom. The field pattern becomes asymmetric due to several reasons, including the mirror asymmetry of the harvesting coil itself since the tuning capacitors and rectifiers are introduced in a non-symmetric fashion. However, if the distance between the harvesting coil and the phantom reaches $5$~cm, then the field again appears unperturbed, as shown in Fig.~\ref{fig:B1_MAP}(d), and closely resembles that in the absence of harvesting coil, Fig.~\ref{fig:B1_MAP}(b). This said the proposed harvesting coil design should not affect the quality of MRI images if placed at sufficient separation from the patient.

The peak voltage during one RF excitation pulse reaches up to $100$~V on the $R=200$~Ohm load. Therefore, the peak power is approximately $50$~W. However, the duration of excitation pulses usually in the range  $100~\mu s$ - $5$~ms, and time repetition (TR) pulse time is typically from $5$~ms to several seconds~\cite{bernstein2004handbook}. Besides, the envelope shape of MR-signal is not a rectangle, as ${\rm sinc}$-function is mostly used. As a result, the estimated average power at each step of the phase encoding cycle is about $100...500$~mW.

%________________________Numerical_simulations________________________
\section{Numerical simulations}

To further detail our studies, we perform a set of numerical simulations with CST Microwave Studio 2020. First, we repeat the experimental protocols starting with harvesting output voltage depends on the distance $z$ shown with solid green curve in Fig.~\ref{fig:Experiment}(c). For such a simulation, we create a model that incorporates a birdcage body coil, homogeneous cylindrical phantom with additional frame resembling the ones used in real experiments, and a detailed model of the harvesting coil, including necessary lumped elements, Fig.~\ref{fig:B1_MAP}(a). As seen from Fig.~\ref{fig:Experiment}(c), numerical simulation results agree well with both Espree and Avanto measurements. However, numerically obtained voltages are greater than experimental ones in the vicinity of isocenter up to $10\%$, which is likely related to the simplified nature of the model does not taking into account realistic body coil design features (such as additional slots with serial capacitors) required for eliminating eddy currents.

\begin{figure*}[t]
    \centering
    \includegraphics[width=17cm]{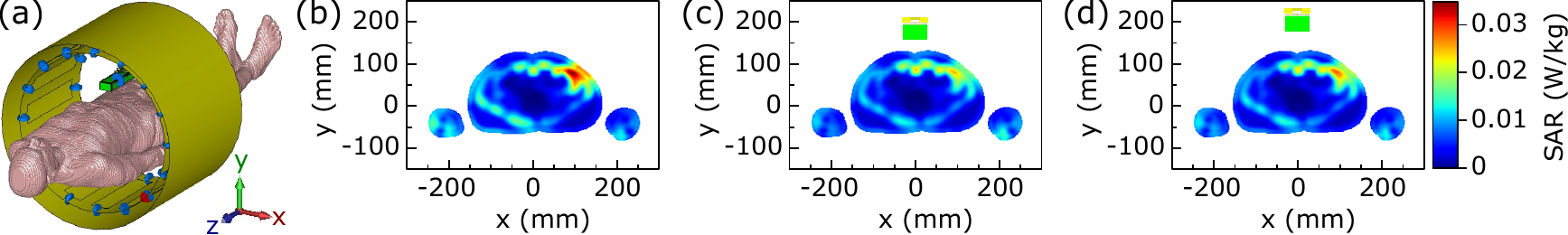}
    \caption{(a) Voxel model of a human used for numerical simulations of the specific absorption rate (SAR), (b-d) SAR images in the case without harvesting coil (b) and with a harvesting coil placed at the height of $1$~cm (c) and $5$~cm (e) above the voxel model.}
    \label{fig:SAR_MAP}
\end{figure*}

Then, we calculate the root mean square (RMS) of the RF field $B_1$ in the absence of the harvesting coil, Fig.~\ref{fig:B1_MAP}(e), and for the harvesting coil located at different distances from the phantom. Fig.~\ref{fig:B1_MAP}(f,g) applying the same numerical model. As seen from the obtained distributions, the field with the harvesting coil placed at the distance $5$~cm closely resembles that in the absence of coil, while for the coil located at $1$~cm the field appears strongly perturbed. These results demonstrate good agreement with the experiment.

The safety of MRI is mostly defined by the SAR, which identifies the amount of RF radiation absorbed by the patient. To verify the patient safety in the presence of the harvesting coil, we perform numerical simulations of SAR. For this, we employ the detailed voxel 3D model of a human body, including internal organs and perform simulations in CST Microwave Studio with the same body and harvesting coil, Fig.~\ref{fig:SAR_MAP}(a). In Fig.~\ref{fig:SAR_MAP}(b), SAR is calculated in the absence of the harvesting coil. The resulting pattern appears asymmetric with a maximum in one hand of the patient. If the harvesting coil placed close to the patient at a distance $1$~cm, that SAR decreases due to the presence of absorption in the coil, highlighting the absence of sufficient field value focusing in the region of patient, Fig.~\ref{fig:SAR_MAP}(c). Finally, if the harvesting coil is located at $5$~cm above the patient as shown in Fig.~\ref{fig:SAR_MAP}(d), then SAR distribution becomes close to the one observed in the absence of coil, still featuring no pronounced increase. Thus, the proposed harvesting coil design appears safe for MRI applications.

%___________________________Discussion_______________________________
\section{Discussion}

We have developed a novel design of harvesting coil for MRI based on the use of two decoupled coils to convert circularly polarized RF field in contrast to linear polarization harvesters developed previously. Such a modification appears natural due to the circular polarization of the field $B_1$ created by the birdcage body coil. The measurements show that it is possible to increase output voltage almost two times with circularly polarized RF-coil. By performing numerical simulations and experiments with real MRI scanners, we demonstrate that the proposed coil is suitable for low power consumption devices, safe for the patient, and does not considerably affect the homogeneity of the RF magnetic field. 

The proposed design can be used as a main or additional power supply for multichannel phased array coils, which are the gold standard in clinical MRI imaging. The amount of channels there can achieve high numbers up to 8 or even higher. Each channel requires power levels of at least~\cite{byron2017rf} 100-300 {\rm mW}. Therefore for the proper functioning of 8 channel array, about 3 {\rm W} would be required, which is unattainable for the current harvesting setup. However, based on our design, it is possible to realize 1-2 channel wireless coil (for example, for dental MRI~\cite{ludwig2016dental}), MRI coil wireless identification, wireless ECG sensors~\cite{baig2013comprehensive}, etc. The directions for further development of the concept include direct testing of harvesting coil use in combination with local coils and sensors.

\section*{Acknowledgments}
This work was supported by the Russian Science Foundation (Project No.21-79-30038).

%\section*{Data Availability}
%The data that support the findings of this study are available from the corresponding author upon reasonable request.

%\bibliography{Harvesting_BibFile}

\end{document}